# Demonstration of whispering-gallery-mode resonant enhancement of optical forces

Yangcheng Li,[1] Alexey V. Maslov,[2] Nicholaos I. Limberopoulos,[3] and Vasily N. Astratov[1]

[1]Department of Physics and Optical Science, Center for Optoelectronics and Optical Communications, University of North Carolina at Charlotte, Charlotte, NC 28223-0001, USA
[2]Department of Radiophysics, University of Nizhny Novgorod, Nizhny Novgorod 603950, Russia
[3]Air Force Research Laboratory, Sensors Directorate, Wright Patterson AFB, Ohio 45433, USA
E-mails: yli63@uncc.edu, astratov@uncc.edu

*Abstract*—We experimentally studied whispering-gallery-modes (WGMs) and demonstrated resonance enhancement of optical forces evanescently exerted on dielectric microspheres. We showed that the resonant light pressure can be used for optical sorting of microparticles with extraordinary uniform resonant properties that is unachievable by conventional sorting techniques.

## I. Introduction

Since the invention of optical tweezers [1] optical forces have been widely used for manipulating microparticles. Weakly pronounced whispering-gallery-mode (WGM) peaks have been observed in the spectra of the light forces exerted on oil droplets [2]. However, these peaks have not been studied in sufficient detail. Recently, we studied optical propulsion of polystyrene microspheres with diameters from 3 to 20 μm using tapered fiber couplers. We observed that some of the spheres having 15-20 μm diameters can be propelled at an extraordinary high velocity [3-6]. Giant optical propulsion velocities of ~450 μm·s$^{-1}$ were recorded for 20 μm polystyrene microspheres in water with only 43 mW guided power [4]. Theoretically, we showed that the peak optical force can approach the limit determined by the total momentum transfer from light to the microparticle [4,7]. These experiments were carried out with an assembly of microspheres interacting with the tapered fiber. Due to the 1-2% size variation of the microspheres and the size-dependent nature of WGM resonant frequencies, we realized a random detuning between the fixed laser emission line and the WGM resonances. Therefore, only a small fraction of microspheres interacted with the tapered fiber resonantly, making our technique ideal for selection of spheres with a much narrower size distribution.

In this work, we demonstrated a method to precisely control the wavelength detuning between the laser emission line and the WGMs of the microspheres. The method is based on trapping individual microspheres by optical tweezers followed by their spectroscopic characterization and by their propulsion with precisely controlled detuning conditions, as shown in the initial results in [8-10]. Spheres sorted by their resonances can be used in coupled resonator optical waveguides and other devices [11].

## II. Experiments and Results

Individual spheres were trapped by optical tweezers and brought to the vicinity of the tapered fiber, as illustrated in Fig. 1(a). While the sphere was on hold, broadband white light was coupled into the fiber and the transmission was recorded by optical spectrum analyzer [12] as illustrated in Fig. 1(b).

With the knowledge of WGMs resonant wavelengths of a given sphere, we reconnected the input fiber to a tunable laser and set the laser emission line to the desirable wavelength detuning relative to a prominent resonant dip in the fiber transmission spectrum. Fig. 1(b) shows a typical transmission spectrum of a 20 μm polystyrene sphere coupled to the tapered fiber in water environment, and the laser emission line was set to one of the resonant dips (zero wavelength detuning). After that, the optical tweezers beam was turned off and the sphere was released, as shown in Fig. 1(c). The released spheres tend to slowly drift away from the fiber, however by switching on the laser beam tuned in resonance with the WGMs the spheres can be radially trapped near the tapered fiber and propelled along the tapered fiber. Fig. 1(d) shows a series of snapshots of a sphere's motion after release with the laser line tuned on resonance as illustrated in Fig. 1(b).

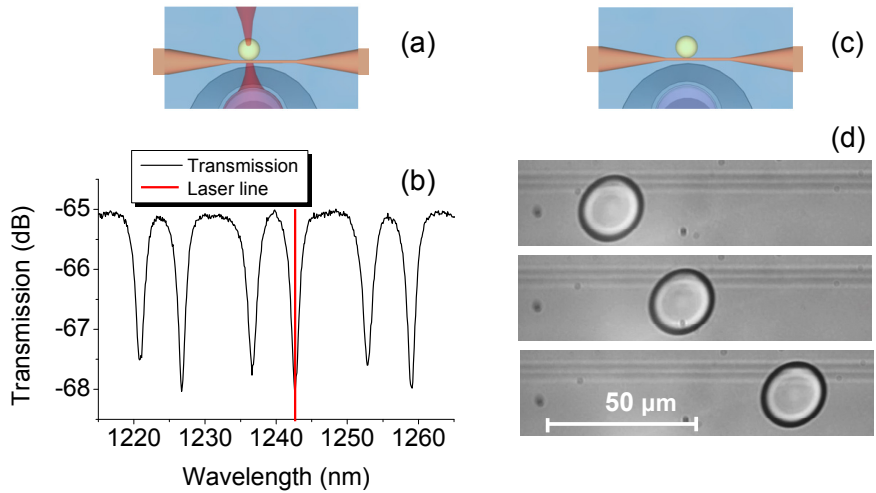

Figure 1. (a) Optical tweezers are used to bring a sphere to the tapered fiber. (b) Measured fiber transmission spectrum. (c) Releasing the sphere by blocking the optical tweezers beam. (d) Snapshots with 100 ms time intervals of sphere being propelled.

The experimental techniques introduced above allowed us to precisely control the laser detuning from the WGMs resonant wavelength. Using a tunable laser, we were able to sweep across an entire resonance and study the spectral dependence of optical forces. Fig. 2 represents the results of such experiments for polystyrene spheres of 10 μm nominal diameters. Multiple points for each value of wavelength detuning represent data obtained using different spheres。

The coupling between tapered fiber and 10 μm spheres in water is weak [4,11], indicated by the broad resonance with the relatively shallow depth seen in Fig. 2. Optical propulsion effects could be observed for the entire range of detuning from -15 nm up to +15 nm. The measured velocities are found to be within 2-5 $\mu m \cdot s^{-1} \cdot mW^{-1}$ range. The spectral peak of the velocity (red points) coincides with the minimum in the fiber transmission spectrum (black curve) and the shape of the peak of the force can be considered a mirror reflection of the shape of transmission spectrum. The resonant enhancement of optical force is weakly pronounced in this case.

In contrast, the coupling to 20 μm spheres is strong, as more than 50% of the optical power can be coupled into WGMs, as can be seen in Fig. 1(b). We found that not only the propulsion velocity, but also the optical attraction of the spheres to the tapered fiber displayed strongly pronounced resonant behavior. If the laser line was detuned from the minimum of the resonant dip by more than 1 nm, the radial trapping of spheres was not observed. In these cases, the spheres were found to slowly drift away from the fiber due to gravity force and background water flux. This behavior indicates that the light-induced radial trapping is not sufficient for retaining the spheres near the tapered fiber in non-resonant cases. We found, however, that when the laser line was tuned towards the center of the resonance, the spheres stayed trapped in the vicinity of the taper. Under these conditions, we were able to observe optical propulsion of such spheres. At near-resonance conditions, a larger portion of the momentum carried by the laser light was transferred to the sphere, contributing to a bigger propulsion force. An extraordinary high velocity of 16 $\mu m \cdot s^{-1} \cdot mW^{-1}$ was observed when the laser hit the resonance dip, exceeding previously recorded high velocity of 10 $\mu m \cdot s^{-1} \cdot mW^{-1}$ [4]. The spectral shape of the velocity peak was also found to be a mirror-image of the shape of the dip in the fiber transmission spectrum.

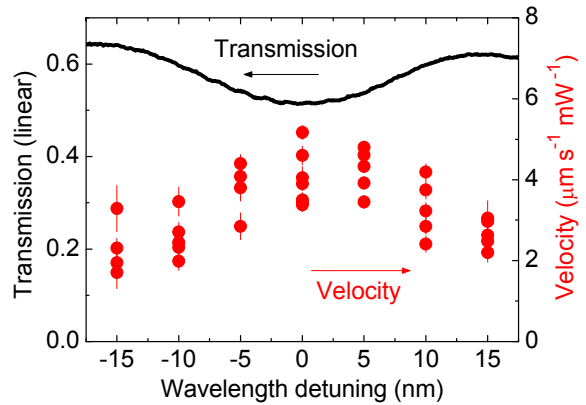

Figure 2. Transmission spectrum and propelling velocities with wavelength detuning for 10 μm diameter spheres.

## III. CONCLUSION

In this work, we presented a method to precisely control the wavelength detuning between the laser emission line and the WGM resonance in the microsphere. By studying the propulsion of dielectric microspheres in water-immersed fiber couplers, we measured the spectral shape of the peak force and established that it correlates with the dips in the fiber transmission spectra. We also showed that in agreement with the theoretical model [4,7] the maximal propulsion velocities ~16 $\mu m \cdot s^{-1} \cdot mW^{-1}$ observed for 20 μm spheres correspond to a complete optomechanical transformation of the light momentum flux. Resonant light pressure effects can be used for optical sorting of microparticles with extraordinary uniform resonant properties ($<10^{-4}$) unachievable by conventional sorting techniques.

ACKNOWLEDGMENT

The authors gratefully acknowledge support from U.S. Army Research Office through Dr. J. T. Prater under Contract No. W911NF-09-1-0450, DURIP W911NF-11-1-0406, DURIP W911NF-12-1-0538. A. Maslov acknowledges support from the Ministry of Education and Science of the Russian Federation through agreement No. 14.B37.21.0892.